\documentclass{pic2012}

\def\runauthor{L. Fields}
\def\shorttitle{Neutrino Cross Sections}
\begin{document}

\title{Neutrino Cross Sections}

\author{Laura Fields}

\address{Northwestern University \\
Evanston, IL, 60201, USA\\
E-mail: laurajfields@gmail.com }

\maketitle

\abstracts{The next generation of neutrino oscillation experiments
  aims to answer many interesting questions, such as whether there is
  CP violation in the neutrino sector and whether sterile neutrinos
  exist.  These experiments will
  require high precision cross-section measurements of various neutrino
  and anti-neutrino interaction channels.  We review results
  and prospects for such measurements from the 
  MiniBooNE, T2K, MINER$\nu$A and ArgoNeuT collaborations.    
}

\section{Introduction} 

The discovery of neutrino oscillations and non-zero neutrino mass has
led to an interesting new set of neutrino-related questions.  For
instance, while the neutrino mixing matrix is increasingly
well-measured, the origin of the dissimilarity between this matrix and
the quark mixing matrix is unknown.  It is also unknown whether there
is CP violation in the neutrino sector, or whether there exist
additional neutrinos (called ``sterile neutrinos'') beyond those that
have already been observed.  In the coming decades, neutrino
physicists will attempt to answer some of these questions through
long- and short-baseline oscillation experiments.  

Generally speaking, neutrino oscillation experiments work by producing
a beam of neutrinos and letting it propagate over a distance of
hundreds of meters 
(in the case of short-baseline experiments) to thousands of kilometers (in the
case of long-baseline experiments) before being detected
in a neutrino detector.  Oscillations are
then inferred by comparing the observed neutrino energy spectrum with
predicted spectra given various oscillation hypotheses.  The predicted
spectra require knowledge of neutrino interaction rates in the
detector.  Oscillation experiments therefore require high
precision measurements of neutrino interaction cross section in
various materials.  This is particularly true of the next generation
of oscillation experiments, which aim to measure much smaller effects
(such as CP violation) than their predecessors.  

The range of neutrino energies of interest to oscillation experiments
varies from roughly 200 MeV to 10 GeV.  Several dedicated neutrino
cross-section experiments, such as MINER$\nu$A and ArgoNeuT, have been
developed to study neutrinos in this energy range.  Additionally,
oscillation detectors, such as MiniBooNE and the T2K near detector
(ND280), can double as cross-section experiments.  Below, we briefly review
the types of neutrino interactions that occur in this energy range,
and then review results from these four experiments.  While many other
experiments have measured neutrino interaction cross sections in recent years, we restrict ourselves to
the experiments mentioned above due to time and length restrictions.

Because all of the detectors discussed here sit in muon neutrino or
muon anti-neutrino beams, all of the reviewed cross-section results are of
muon neutrino or muon anti-neutrino interactions, although we note that some
experiments also plan to make measurements of the
small electron-neutrino component of their beam.
Additionally, we review only neutrino interactions with nuclei.  While
neutrino interactions with electrons do occur, and
will be probed by experiments such as MINER$\nu$A, an understanding of
neutrino-nucleus interactions is most crucial to oscillation
experiments and is thus the focus of this summary. 

\section{Overview of Neutrino Interactions} 

When a neutrino interacts with a nucleus within a particle detector, it can exchange either a
charged $W$ or a neutral $Z$ boson with the nucleus; these
interactions are called ``charged-current'' or ``neutral-current'',
respectively.  Each of these interactions and their relevance to
oscillation experiments is described below.

\subsection{Neutral-Current Interactions}
 Neutral current interactions are characterized by the presence of a neutrino in both
the initial and final states.  Because neutrinos are 
invisible to particle detectors, it is difficult to reconstruct the
energy of the incoming neutrino in neutral current interactions, and
they are generally not used as signal processes in oscillation
experiments.  Understanding their cross sections is still important, as
neutral current interactions can be backgrounds to oscillation
signals.  In particular, the channel often referred to as ``neutral
current $\pi^0$ production'' ($\nu N\rightarrow\nu\pi^0X$) often appears in a detector
as a single $\pi^0$, which can fake a signal electron in a $\nu_e$
appearance measurement.

\subsection{Charged-Current Interactions}
Charged-current neutrino-nucleus interactions are characterized by the
presence of a charged lepton (matching the flavor of the initial state
neutrino) in the final state.  The absence of a neutrino in the final
state makes energy reconstruction of these interactions considerably
simpler than for neutral-current interactions, and charged-current
interactions are often used as signal channels in oscillation
measurements.  Of particular interest is the channel known as
``charged-current quasi-elastic (CCQE)'':
\begin{eqnarray}
\nu_\mu n \rightarrow \mu p \\
\bar{\nu}_\mu p \rightarrow \bar{\mu}n.
\end{eqnarray}
If one assumes that the initial state nucleon is at rest, the
kinematics of this interaction can be completely reconstructed using
muon kinematics only.  

\begin{figure}
\begin{center}
\includegraphics[scale=0.4]{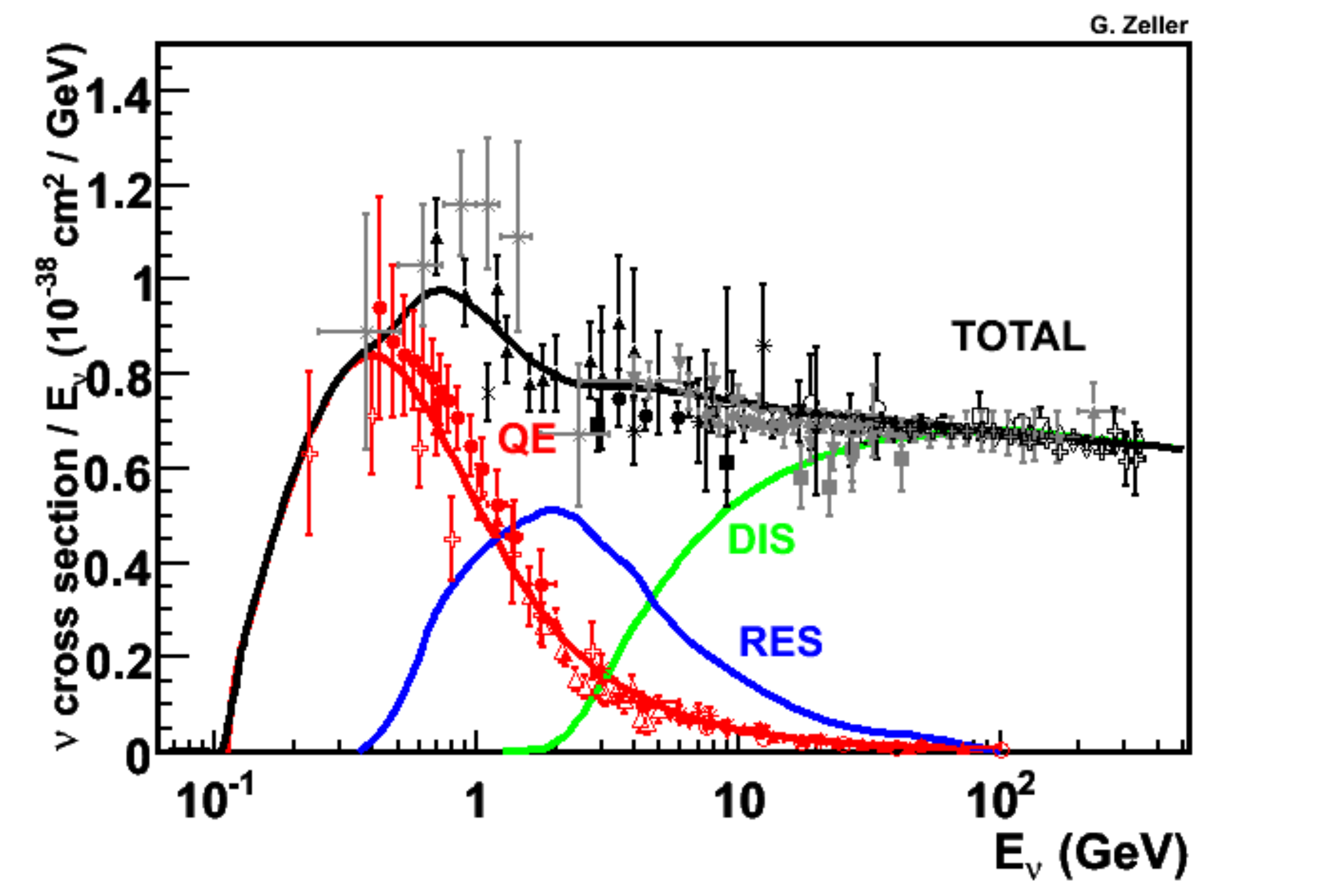}
\end{center}
\caption[*]{Summary of the current knowledge of $\nu_\mu$ charged-current
 cross sections.  Plot courtesy of G. Zeller~\cite{FZ}.}
\label{fig:nuchannels}
\end{figure}

Figure~\ref{fig:nuchannels} summarizes our
current knowledge of charged-current neutrino scattering
cross sections.  It shows that quasi-elastic scattering is the
dominate charged-current cross section below 1 GeV, but at higher
energies other processes such as resonant pion production and deep inelastic scattering
become dominant.  The resonant pion channel, particularly in
cases where the pion has too little momentum to be seen in a detector
or is absorbed in the target nucleus, is a common background to quasi-elastic scattering signals.  

\subsection{Nuclear Effects}
Interpretation of both oscillation and cross-section measurements is
complicated by the fact that modern neutrino detectors are
usually made of heavy nuclei.  Neutrino interactions with nucleons
bound within a nucleus differ from those with free nucleons in a
number of ways.
\begin{itemize}
\item The interaction cross section for bound nucleons is reduced by
  effects such as Pauli blocking,
  particularly in the kinematic region where the momentum transferred
  from the neutrino to the nucleon is low.
\item The initial-state nucleon has a non-zero Fermi-momentum, which
  effectively smears kinematic reconstruction that typically assumes
  a nucleon at rest.
\item Final state particles can undergo interactions as they traverse
  the nucleon.  These interactions can result in significantly
  different final-states than were produced by the primary
  interaction.
\item Neutrinos can interact with multi-nucleon bound states within the nucleus.  Whether
 such interactions contribute significantly to the total interaction
 cross section is currently unclear.  If they do, this could have
 significant implications for oscillation experiments, as they may
 be difficult to distinguish from single-nucleon final-states in many detectors.
\end{itemize}
All of these effects may vary depending on the type of
nucleus involved in the interaction.  Untangling these various effects
and understanding how they vary with different nuclei is a primary focus
of neutrino cross-section experiments.  

\section{MiniBooNE}

\begin{figure}
\begin{center}
\includegraphics[scale=0.6]{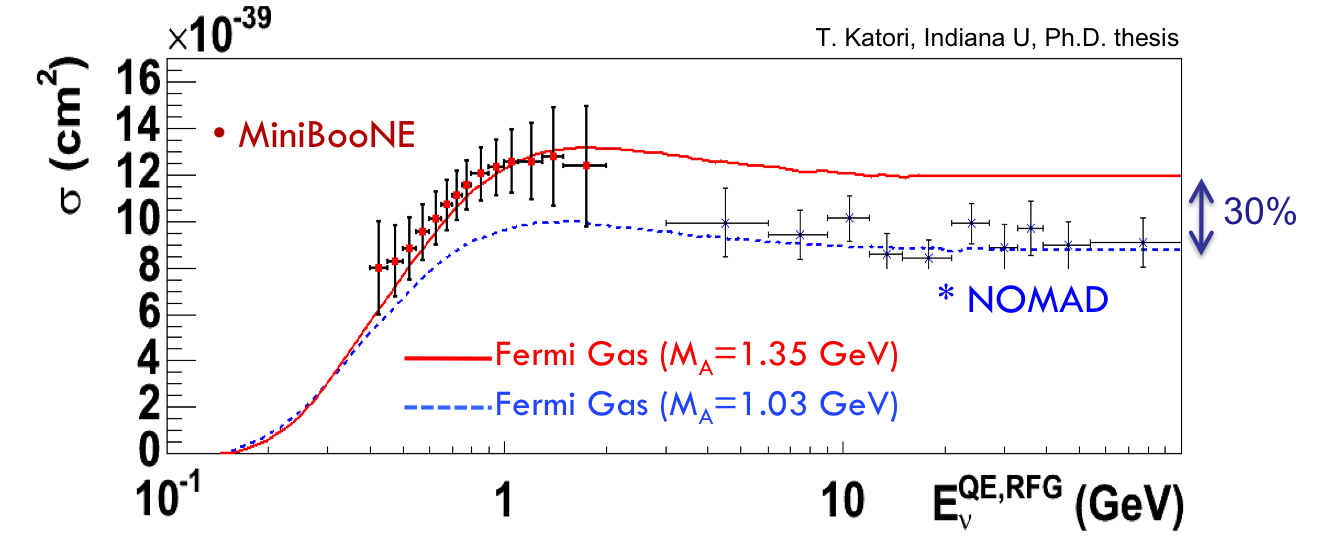}
\end{center}
\caption[*]{MiniBooNE $\nu_\mu$ charged-current quasi-elastic
 cross-section measurements~\cite{AguilarArevalo:2010zc,katori_thesis} as a function of neutrino energy,
 compared with results from the NOMAD experiment~\cite{Lyubushkin:2008pe}.  Plot courtesy of T. Katori.}
\label{fig:minibooneccqe}
\end{figure}

Although most famous for short-baseline oscillation measurements, the
MiniBooNE collaboration has also been a prolific source of
cross-section measurements.  Located in the Booster neutrino beam at
Fermilab, the MiniBooNE detector is a 12.2 diameter sphere of mineral
oil surrounded by 1280 photo-multiplier tubes, which detect both
Cerenkov and scintillation light produced by neutrino
interactions within the mineral oil.  The detector sees an average
neutrino energy of approximately 700 MeV.

One of MiniBooNE's first cross-section measurement was of the
$\nu_\mu$ charged-current quasi-elastic channel~\cite{AguilarArevalo:2010zc}, shown in
figure~\ref{fig:minibooneccqe}.  One parameter measured by MiniBooNE is
``$M_A$'' -- the mass term in the axial form factor.  While many
previous experiments have measured $M_A$ to be near unity~\cite{Bernard:2001rs}, the
MiniBooNE data prefer a higher value of $\mathrm{M_A} = 1.35\pm0.17$ GeV$/$c$^2$.  The origin of this discrepancy is
currently unclear, but one hypothesis is that the MiniBooNE
measurement includes quasi-elastic-like interactions that involve
neutrino interactions with multi-nucleon states.  This topic has
become a rich field of theoretical speculation (see for
example~\cite{Martini:2009uj,Nieves:2011pp,Bodek:2011ps}),
and investigating the multi-nucleon contribution to the
quasi-elastic cross section is likely to be a major focus of future
quasi-elastic measurements.

Following this very interesting quasi-elastic result, MiniBooNE has
gone on to produce eight neutrino cross-section
papers~\cite{AguilarArevalo:2010zc,AguilarArevalo:2010bm,AguilarArevalo:2010xt,AguilarArevalo:2010cx,AguilarArevalo:2009ww,AguilarArevalo:2009eb,AguilarArevalo:2008xs,AguilarArevalo:2007ab},
measuring most of the interactions present in their $\nu_\mu$
beam.  The collaboration is currently working on similar measurements
using their $\bar\nu_\mu$-enhanced beam and on reassessing all of
their measurements in light of the new possibility of neutrino
interactions with multi-nucleon states.  

\section{T2K}

The off-axis T2K near detector~\cite{T2K}, known as ND280, serves as the near
companion to the Super-K detector for T2K's long baseline studies.  It
is intended to measure neutrino interaction cross sections and other
inputs needed for oscillation results.  It is composed of several
detectors, including an upstream $\pi^0$ detector made of
scintillator and water, fine grained detectors composed of 1 cm
square bars of scintillator, and time projection chambers.  ND280 is
somewhat unusual among neutrino cross-section detectors in that it
includes a magnet, creating a 0.2 T magnetic field in the inner
detectors.  The detector sits about 2.5 degrees off axis in a neutrino
beam generated at the J-PARC accelerator facility and experiences a
beam with a mean neutrino energy of around 700 MeV.

T2K's first cross-section measurement is an inclusive measurement of
all charged-current channels~\cite{T2KCC}.  The muon momentum and angular
distributions of this sample are shown in Figure~\ref{fig:t2kcc}.
They find a total flux-averaged cross section of
$(6.93\pm0.13\pm0.085)\times10^{-39}$ cm$^2/$nucleon, which is
consistent with generator predictions that have been tuned with prior data  They have further divided
this sample into quasi-elastic and non-quasi-elastic enriched
samples~\cite{T2KCCQE}, and have fit these samples to extract cross-section
parameters that are used in their oscillation measurements.  The
collaboration will begin producing an array of complete cross-section
measurements soon, and may also make cross-section measurements using 

\begin{figure}
\begin{center}
\includegraphics[scale=0.4]{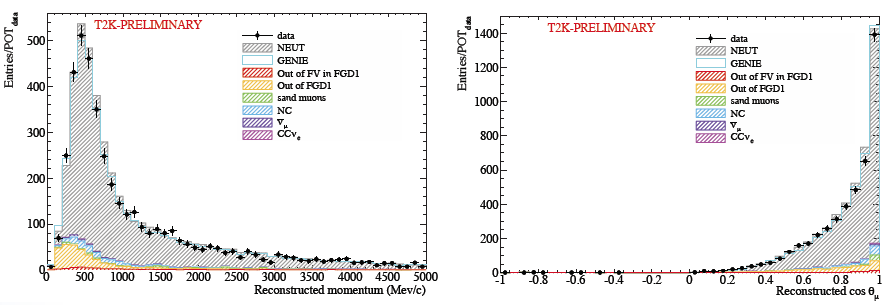}
\end{center}
\caption[*]{Muon momentum and angular distributions in T2K's
 charged-current quasi-elastic sample.}
\label{fig:t2kcc}
\end{figure}

\section{MINER$\nu$A}

The MINER$\nu$A experiment is a dedicated neutrino cross section 
experiment located in the NuMI beamline at Fermilab.  It contains a
large volume of plastic scintillator interspersed
with various inactive materials.  The upstream portion of the detector
includes modules of inactive carbon, iron, lead, water and helium, which
will be used to compare neutrino interaction cross sections on
different nuclei. The detector sits just upstream of the MINOS near
detector, which functions as MINER$\nu$A's muon spectrometer.

\begin{figure}
\begin{center}
\includegraphics[scale=0.4]{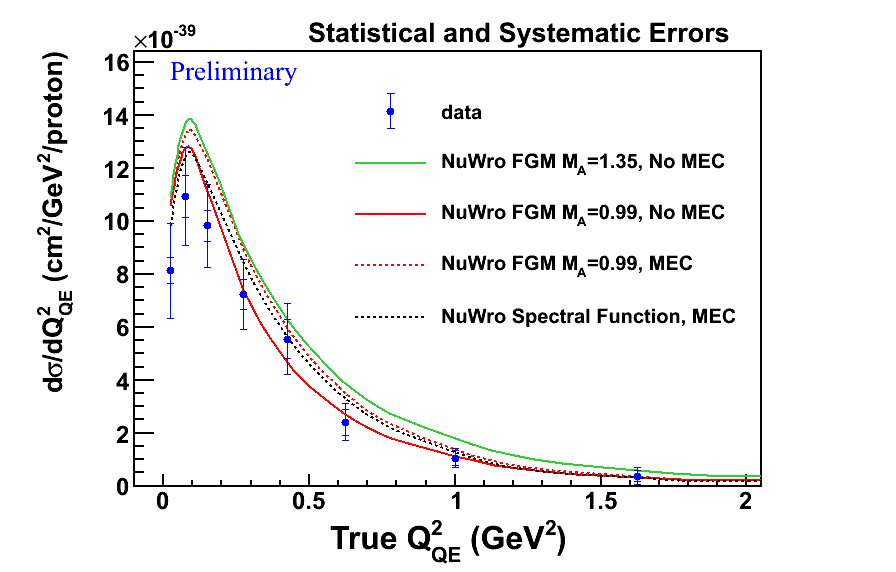}
\end{center}
\caption[*]{$\bar{\nu}_\mu$ differential cross sections measured in
 MINER$\nu$A, compared with various models.  The models labeled
 ``MEC'' include interactions on multi-nucleon~\cite{Bodek:2011ps}.
 bound states.}
\label{fig:minervaccqe}
\end{figure}

The MINER$\nu$A collaboration has recently presented its first
cross section measurement, a flux-integrated $\bar{\nu}_\mu$
quasi-elastic  differential
cross section with respect to $Q^2$ (where $Q$ is the 4-momentum
transferred from the initial-state neutrino to the final-state
nucleon), shown in Figure~\ref{fig:minervaccqe}.  This differential
cross section has also been compared with various models, including
those that include a multi-nucleon enhancement to the
quasi-elastic cross section.  While the data do not yet
have sufficient statistical power to rule out any model, the
collaboration has more than a factor of three times more data on tape as
well as plans to significantly improve systematic uncertainties,
making it likely that MINER$\nu$A will have a more definitive
statement on the presence of significant multi-nucleon contributions in the near future.

The MINER$\nu$A collaboration has also recently presented preliminary
results in a variety of other channels, including several pion channels
and $\nu_\mu$ quasi-elastic interactions in both scintillator and the nuclear
targets.  The experiment expects to publish its first papers using its
initial low energy run in the next year, and will also begin a medium
energy run using the NO$\nu$A configuration of the NuMI beam in 2013.  

\section{ArgoNeuT}

The ArgoNeuT detector is a 170 liter liquid Argon TPC that sat
between MINER$\nu$A and the MINOS near detector during several months
of 2009 and 2010.  As the first liquid Argon detector to reside in a
low-energy neutrino beam, ArgoNeuT is an important step in the
development of kiloton-scale liquid Argon detectors.  Liquid Argon
TPCs offer resolution comparable to bubble
chambers, which gives it unprecedented ability to visualize
interactions.  Of particular interest is ArgoNeuT's ability to reconstruct activity
near the interaction vertex, which is a sensitive probe of nuclear
effects.   ArgoNeuT's first
cross section measurement is of the charged-current inclusive
cross section, where they find $\sigma/E_\nu=(7.3\pm1.2)\times10^{-39}$
cm$^2$/GeV/nucleon at $\langle E_\nu\rangle=4.3$ GeV, which is consistent
with Monte Carlo predictions~\cite{Anderson:2011ce}.  The
collaboration is now working towards measurement of exclusive
channels, including charged-current quasi-elastic.

\begin{figure}
\begin{center}
\includegraphics[scale=1.0]{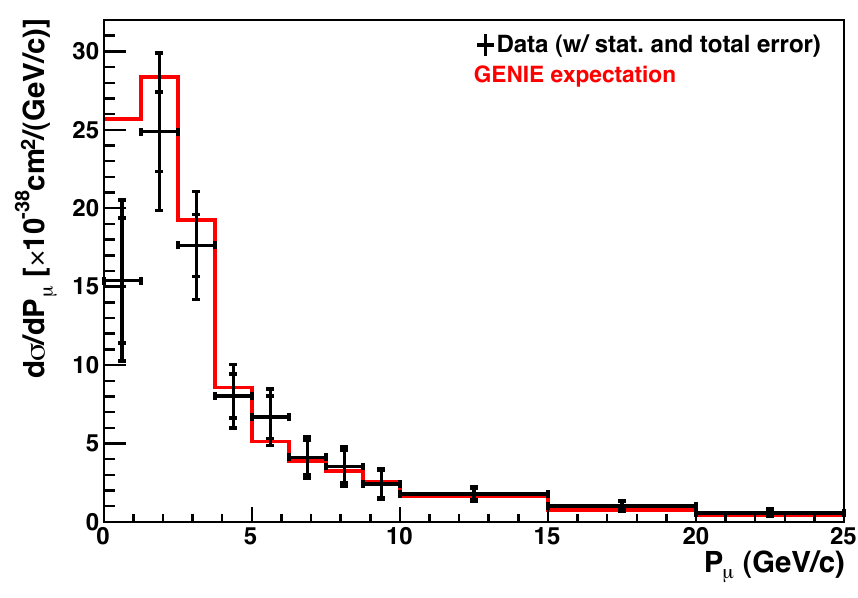}
\end{center}
\caption[*]{Muon momentum spectrum in ArgoNeuT's $\nu_\mu$ charged-current
  inclusive sample.}
\label{fig:t2kcc}
\end{figure}

\section{Conclusion}

The field of neutrino cross section measurements is both rich and
important.  The next generation of neutrino oscillation experiments
will require a wide array of well-measured neutrino interaction
cross sections.  Because neutrinos interact with nucleons bound within
the nuclei of neutrino detectors, making such measurements also gives
us a unique glimpse at nuclear structure.  Several experiments which
have measured cross sections, such as MiniBooNE and ArgoNeuT, have
completed data-taking and are currently making their final
measurements.  Others, such as MINER$\nu$A and T2K's ND280, are in the
midst of their data runs and have begun producing preliminary results
that show much promise for the future.  In the next few years, new
experiments such as MicroBooNE, ICARUS and the NOvA near detector will come
online, ensuring that the field of neutrino cross section measurements
will be active for many years to come.

\section*{Acknowledgements} 

This work was supported by the U.S. Department of Energy.  The author would
like to thank the MiniBooNE, T2K, MINER$\nu$A and ArgoNeuT
collaborations, with particular gratitude to D. Harris, K. Mahn,
K. McFarland, M. Soderberg and G. Zeller for their assistance in
compiling this material


\end{document}